\documentclass[conference]{IEEEtran}
\IEEEoverridecommandlockouts
\usepackage{cite}
\usepackage{amsmath,amssymb,amsfonts}
\usepackage{algorithmic}
\usepackage{graphicx}
\usepackage{textcomp}
\usepackage{booktabs}
\usepackage{diagbox} 
\usepackage{multirow}
\usepackage{siunitx}
\usepackage[table]{xcolor}
\def\BibTeX{{\rm B\kern-.05em{\sc i\kern-.025em b}\kern-.08em
    T\kern-.1667em\lower.7ex\hbox{E}\kern-.125emX}}
\begin{document}
\newcommand{\red}[1]{\textcolor{red}{#1}}

\title{Biodenoising: Animal Vocalization Denoising without Access to Clean Data}

\author{\IEEEauthorblockN{Marius Miron, Sara Keen, Jen-Yu Liu, Benjamin Hoffman, Masato Hagiwara, \\ Olivier Pietquin, Felix Effenberger, Maddie Cusimano}
\IEEEauthorblockA{\textit{Earth Species Project} \\}
\and
}

\maketitle

\begin{abstract}
Animal vocalization denoising is a task similar to human speech enhancement, which is relatively well-studied. 
In contrast to the latter, it comprises a higher diversity of sound production mechanisms and recording environments, and this higher diversity is a challenge for existing models.
Adding to the challenge and in contrast to speech, we lack large and diverse datasets comprising clean vocalizations. 
As a solution we use as training data pseudo-clean targets, i.e. pre-denoised vocalizations, and segments of background noise without a vocalization. 
We propose a train set derived from bioacoustics datasets and repositories representing diverse species, acoustic environments, geographic regions.
Additionally, we introduce a non-overlapping benchmark set comprising clean vocalizations from different taxa and noise samples.
We show that that denoising models (demucs, CleanUNet) trained on pseudo-clean targets obtained with speech enhancement models achieve competitive results on the benchmarking set. We publish data, code, libraries, and demos at https://mariusmiron.com/research/biodenoising.
\end{abstract}


\section{Introduction}\label{sec:intro}

Animal vocalization denoising is the task of removing noise from soundscapes or focal audio recordings of animals recorded with microphones or hydrophones.
Here, the signal is the focal animal vocalization and the noise is all other sound sources.
The nature of the noise may be anthropogenic (e.g. engine sounds, boats, explosions, surveys, factories), natural (e.g. wind, tides, waves, but also other animal vocalizations not generated by the focal species) 
or of other mechanical (e.g. clicks, pops, crackles, mechanical) or electrical (e.g. hum, buzz, static) origins~\cite{fournet2018more}. 
%
Note that in this work we do not aim at separating between overlapping animal sounds~\cite{lin2020source,denton2022improving}, i.e. solving the cocktail party problem for mixtures of animal sounds, a task analogous to human speech separation.



Recordings of animal vocalizations are used to study animal communication and animal behavior~\cite{bradbury1998principles}, and to monitor biodiversity~\cite{stowell2022computational}. However, 
noise presents a large challenge for biologists aiming to evaluate acoustic differences among signals or to automate the measurement of acoustic variables~\cite{fischer2013bioacoustic}. More, playback experiments, which are a crucial tool for studying animal communication,  require good quality, possibly noise-free recordings~\cite{fournet2018more}.

In this paper and in contrast to the speech domain we aim at denoising animal vocalizations without having access to clean data which is lacking for most species.
%
%
%
%
To achieve that we follow a two-step process. First, we use \textbf{speech enhancement models}, demucs dns48~\cite{defossez2020real} or CleanUNet~\cite{kong2022speech}, to denoise noisy animal vocalizations, obtaining pseudo-clean targets. Second, we re-train these models on synthetic mixtures of noise and pseudo-clean targets. 
\noindent\textbf{Contributions. (1)} We introduce a diverse handcrafted benchmark dataset (Section \ref{ssec:benchmark}) containing clean vocalizations and noise. For training, we propose a large noisy training set (Section \ref{ssec:train}) derived from existing labeled bioacoustics datasets from which we extract excerpts of noisy vocalizations, as well as background noise without a vocalization. 
\textbf{(2)} 
We propose a denoising method when solely noisy vocalizations are available by leveraging existing speech enhancement models. 
Surprisingly, we found that the pseudo-clean targets obtained with speech enhancement may represent good training data despite their low quality.
Our method shows promising generalization performance for the non-overlapping benchmarking set when compared to `noisereduce'~\cite{sainburg2020finding}. 

\section{Relation with previous work}
\label{sec:relation}


Speech enhancement has benefited from publicly-available large clean datasets~\cite{panayotov2015librispeech,wichern2019wham,fonseca2021fsd50k} and benchmarks~\cite{reddy2020interspeech}. This is not the case for animal domain. In this research we formalize the denoising task for the animal domain by introducing a large training dataset and a non-overlapping benchmarking set.

The lack of clean training data presents a challenge for model development, similar to unsupervised speech enhancement~\cite{wisdom2020unsupervised,ito2023audio,tzinis2022remixit,fujimura2023analysis,hao2023neural,saijo2023self}.
One way to approach this task is training on mixtures of noisy targets and noise, known as `noisy target training'~\cite{fujimura2023analysis}. 
Here we prove that leveraging existing speech enhancement models to obtain pseudo-clean targets is better than using noisy targets. 

To our best knowledge, besides generic denoising methods such as `noisereduce'~\cite{sainburg2020finding}, most of the methods are not evaluated across multiple species and environments~\cite{bergler2020orca,barnhill2024animal,kumar2024vision,stowell2015denoising}.
%
Instead, we train a model that is tested for generalization capabilities on a non-overlapping benchmark. 
We take inspiration in recent work showing that self-supervised speech models offer good priors for various bioacoustics classification tasks \cite{sarkar2024utility,cauzinilleinvestigating,kloots2024exploring} and that classifier models trained on bird signals perform well when used for other taxa~\cite{ghani2023global}. 
These recent findings suggest that cross-taxa similarities in signal characteristics and recording environments may allow for generalization.

Although previous work applying unsupervised speech separation to bird datasets has shown promise~\cite{denton2022improving}, this was not the case for our datasets.
While we tried to adapt this approach to denoising, the heuristics necessary to classify those sources as either vocalization or noise, e.g. using a classifier or a domain-aware event detector, did not generalize well and we decided not to pursue this approach. 

\section{Datasets}
\label{sec:datasets}
To establish the denoising task in the animal domain we introduce a train (Section \ref{ssec:train}) and a benchmarking dataset (Section \ref{ssec:benchmark}).
To facilitate downloading and processing these data, and to allow for open research and reproducibility we develop a Python library \texttt{biodenoising-datasets}~\cite{code-datasets}.
The datasets are generated at two target sample rates: 16 and 48 kHz. The original sample rates vary between 44.1kHz and 96kHz for vocalizations, and 16 to 96kHz for noise. 
We resample the vocalizations and time-scale the noise similarly to the procedures used in creating the DNS Challenge dataset~\cite{reddy2020interspeech}.

\subsection{Train set}\label{ssec:train}

To devise a large and diverse training set, we rely on openly available noisy bioacoustics datasets containing isolated calls or soundscapes annotated with vocalizations start and end times. For the soundscapes, we crop the audio frames corresponding to these annotations to derive a noisy vocalization set. The remaining frames are used to create excerpts for the noise class. 

We include two dataset of isolated sound samples:
Animal Sound Archive at the Museum für Naturkunde Berlin reference set~\cite{asa}, and
Watkins Marine Mammal Sound Database~\cite{watkins}.
We add to that bioacoustics datasets used in classification: 
Anuran Kaggle~\cite{akbal2023explainable}, Gelada vocalizations~\cite{gustison2012derived}, Macaques coo calls~\cite{fukushima2015distributed},
and detection:
Thyolo, Wydah, and Lemurs from~\cite{dufourq2022passive}, a labeled subset of Xeno-Canto and the Gibbons dataset from~\cite{jeantet2023improving}, the orca detection datasets from Orcalab (SKRW)\cite{poupard2020massive} and Orchive~\cite{ness2013orchive}, the humpback whales datasets from~\cite{fournet2018more,poupard2020massive}, the marine recordings from Sabiod~\cite{glotin2014soundscape}. We discard the noise samples from Xeno-canto, Lemurs, Thyolo, Wydah, Gibbons, the SKRW Orcalab sites, Orchive because they may contain other unannotated species.

To the noise set, we add underwater passive acoustic monitoring recordings from NOAA's Sanctsound project~\cite{noaa} (the sites `ci',`fk',`gr',`hi',`mb',`oc',`pm',`sb') and MBARI's MARS site~\cite{mars} which were automatically classified as not containing animal vocalizations by the respective institutions in the accompanying metadata. We also add ship engine noise from ShipsEar~\cite{santos2016shipsear}, Deepship~\cite{irfan2021deepship} and Orcalab~\cite{poupard2020massive}. For terrestrial noise, we complement our dataset with FSD50K~\cite{fonseca2021fsd50k} filtering out human, animal, and musical sounds.

The train set comprises 1770826 noise excerpts of 4 seconds each. In addition, we have 120688 noisy vocalizations of the same duration, distributed as follows: Anuran 1604, ASA Berlin 10280, Gibbons 2331,Lemurs 3179, Macaques 7285, Orcasound 5617, Orchive 138, Sabiod 1782,  Humpback whales 27388, Thyolo 2531, Watkins 16309, Whydah 1169, Xeno-canto 12665. We publish a table listing the training data sources at the accompanying web page \cite{page}.

\subsection{Benchmark set}
\label{ssec:benchmark}
Towards formalizing animal vocalization denoising task we introduce a test set comprising mixtures of clean vocalizations and noise. We manually selected clean animal vocalizations from Freesound~\cite{freesound}, NOAA~\cite{noaa}, and biology repositories. 
We include vocalizations from the following animals: quail, chicken, lion, pig, horse, seagull, nightingale, gorilla, hedgehog, cat, frog, macaques, peacock, grey seal, elephant, blackbird, marmoset, lyre bird, dolphin, hornbill, minke whale, fin whale, humpback whale. For noise, we manually downloaded and cropped noise samples from Freesound or passive acoustic monitoring repositories containing in general 80-95\% noise~\cite{noaa}. 
Based on the metadata of the files, the underwater noises are engines, rain, and hydrophone recordings capturing waves. The terrestrial noises have tropical forest ambience, rain, and city ambience (cars, trains). 

We programatically create mixtures by pairing vocalizations of noise at random Signal-to-Noise Ratios (SNR) from an uniform distribution between -5 and 10 dB (2.8 average SNR). To ensure reproducibility, we start with a fixed seed that controls the SNR of the mixtures. The samples are between 1 to 60 seconds long (20.14 seconds on average). We split the vocalizations and noises into two lists: underwater (11 vocalizations and 26 noises) and terrestrial (51 vocalizations and 20 noises). For each separate case, we sort the vocalizations and the noise samples and pair them in the order of their duration e.g. matching the longest calls with longest noises. 

To control for the influence of SNR on the quality of denoising, we build a fixed-SNR version of the dataset at the following levels: -5, 0, 5, 10 dB (62 files each). Additionally, we introduce a large version of the dataset by creating all possible combinations for each of the two scenarios: underwater and terrestrial.

To test generalization of the methods and in contrast with common speech enhancement methods that offer train and test splits, we built the benchmark so it does not overlap with the train set in terms of data sources. To that extent, the test domain is unknown when training the model. In terms of species overlap, the benchmark contains three recordings of whales (humpback, minke, fin), species also present in noisy recordings in two subsets of the train set. In terms of genus overlap, there exists a single recording of a Barbary macaque in the test dataset and a small subset of Japanese macaques coo calls in the train set. Despite that the recordings differ in terms of vocalization type, geographic location, type and gear of recording (focal vs. soundscape). The benchmark dataset is made available~\cite{benchmark}.

\section{Methods}\label{sec:methods}

Given the single-channel signal $x \in \mathbb{R}, x=s+n$, that is a mixture between a clean animal vocalization $s$ and noise $n$, our task is to find $\hat{s}$, an estimation of $s$. 
The goal is to compute $\hat{s}=f(x,\Phi)$ by learning $\Phi$, the parameters of the model. When the clean signals $s$ are available, $\Phi$ are learned by minimizing the prediction error $\mathcal{L}=\mathbb{E}[\mathcal{D}(f(x,\Phi),s)]$, where $\mathcal{D}$ is a distance function, such as the L1 or L2.

In our case, the clean signals $s$ are not available. One proposed solution is to add extra noise to the already-noisy input $x$, with $x$ then becoming the `noisy target' to estimate~\cite{bergler2020orca}. The noisier mixtures 
$\overline{x}=x+\overline{n}$ are created by adding extra noise $\overline{n}$ and then $x$ is estimated. 

In this paper, instead of using $x$ as noisy targets~\cite{bergler2020orca,barnhill2024animal}, we propose to use pseudo-clean targets $s^{\prime}$ computed by $f^{\prime}(x,\Phi^{\prime})$, a denoising method such as an already existing speech enhancement model. 
Here we test three pre-denoising methods $f^{\prime}(x,\Phi^{\prime})$:  noisereduce~\cite{sainburg2020finding}, and the speech enhancement models demucs~\cite{defossez2020real}, and CleanUNet~\cite{kong2022speech}. 
We follow the approaches taken in~\cite{defossez2020real,kong2022speech} and we do not estimate the noise. 

We explore whether the parameters of the speech denoising method $\Phi^{\prime}$ are good priors when learning $\Phi$. This formulation is similar to the teacher-student domain adaptation approach~\cite{tzinis2022remixit,saijo2023self} for a single update. \textbf{Our research hypothesis} is that the estimations $\hat{s}^{\prime}$ obtained from training with pseudo-clean targets $s^{\prime}$ have better quality than the pseudo-clean targets themselves and than the estimations from the `noisy target' baseline~\cite{bergler2020orca}, for unseen contexts.

\noindent\textbf{Time-scaled pseudo-clean targets.} %
Previous research points out that there is a correlation between the body size of an animal and the pitch and tempo of vocalizations~\cite{bradbury1998principles}, i.e. smaller animals vocalize with higher pitch at faster tempos. Similar patterns may be reproduced at different time and frequency scales~\cite{page}. Similarly to ~\cite{kloots2024exploring} and towards obtaining better results from speech enhancement models  $f^{\prime}(x,\Phi^{\prime})$, we scale the input $x$ on the time axis, denoted time-scaling, a computationally inexpensive and biologically plausible transformation described in~\cite{guzhov2022audioclip}. Because the scale factor that produces the best results varies for each dataset, we programatically slow down the audio playing it at speeds reduced by factors of 2, 3, and 4. Then, each estimation $\hat{s}^{\prime}$ is scaled back to the original speed. To obtain a single signal, we average all the estimations. We observed that the speech enhancement models sometimes output silence as the estimated target $s^\prime$, and that time-scaling resulted in more useable, non-silent pseudo-clean signals.

\noindent\textbf{Post-processing of the pseudo-clean targets.} %
While reduced with time-scaling, some of the estimations $s^{\prime}$ of the speech enhancement models trained with an L1 loss~\cite{defossez2020real,kong2022speech} may be silent for sounds that are not in the speech frequency range. We filter the silent segments using a peak-detection algorithm, scipy's `find\_peaks', applied to the RMS curves of the signals, similarly to ~\cite{denton2022improving}. To create training excerpts of same duration, we take windows of length $T$ around each of these peaks and we write them as separate training instances~\cite{denton2022improving}. For shorter duration signals, we either zero-pad them to left and right or we repeat them until they reach the length $T$. 
Because speech enhancement models we use \cite{defossez2020real} remove reverberation, we randomly convolve 50\% of the pseudo-clean targets $s^{\prime}$ with the room impulse responses from the DNS Challenge~\cite{reddy2020interspeech}.

\section{Experiments}
\label{sec:experiments}
We compare two main methods, noisereduce~\cite{sainburg2020finding} and a neural network architecture, demucs dns48~\cite{defossez2020real} previously used in speech enhancement. The neural network is trained on two different types of data: \textbf{noisy} targets~\cite{bergler2020orca} or \textbf{pseudo-clean} targets. The pseudo-clean targets are computed either with noisereduce~\cite{sainburg2020finding} or demucs~\cite{defossez2020real}. We also test the impact of time-scaling the inputs to demucs as described in Section \ref{sec:methods}.
%
We release the code~\cite{code} and sound examples~\cite{page} for all the experiments. 

\subsection{Experimental setup}
\label{ssec:setup}
The models operate at 16kHz. We programatically generate mixtures of audio samples of the length of $T=4$ seconds by randomly pairing noisy~\cite{bergler2020orca} or pseudo-clean targets with noise. 
Note that filtering the almost silent pseudo-clean targets obtained with the speech models we are left with 15000 training excerpts and with 30000 when using time-scaling. 
As data augmentations, we use time-scaling as in~\cite{guzhov2022audioclip} with a random factor between $-4$ and $4$, higher than it was used in the human domain to account for the variety of pitches and time scales in the animal domain. We found that uniformly sampling the noisy set and down-sampling large vocalizations sets are important to generalization.

We train using the mini-batch gradient descent algorithm with an Adam with a cyclic learning rate scheduler that was observed to improve convergence speed and generalization performance on unseen data.
%
The demucs models are trained with an L1 loss on the waveform while the CleanUNet we use the L1 loss and a multi-resolution STFT loss~\cite{hao2023neural}. To avoid overfitting and for numerical stability we clip the gradients within the range $-30$ and $30$. For demucs we use a batch size of 32, while for CleanUNet we lower it to 16 due to GPU memory constraints.
We train all models for the same number of steps.  
All the experiments were run on a computing instance with a V100 GPU.
For noisereduce we use the default options: non-stationary noise estimation, without noise as an input, since our test set does not consider parallel noise samples. 
%

\begin{table*}[ht!]
  \caption{SI-SDR and SI-SDRi in \SI{}{dB} reported as median with 95\% confidence intervals across the excerpts from various test subsets. We compare \textbf{noisereduce} with \textbf{demucs dns48} trained on four different targets. Higher values for medians indicate better results.}
  \label{tab:res1}
\footnotesize

\addtolength{\tabcolsep}{-0.35em}
\begin{tabular}{@{}lll|ll|ll|ll|ll@{}} 
& \multicolumn{8}{c}{\textbf{demucs dns48~\cite{defossez2020real} architecture trained with:} } & \multicolumn{2}{|c}{\textbf{noisereduce~\cite{sainburg2020finding}}} 
\\\cmidrule(lr){2-11}

\rowcolor[gray]{.9} \cellcolor{white}{}& \multicolumn{2}{c}{\textbf{ noisy targets~\cite{bergler2020orca}}} & \multicolumn{6}{c|}{\textbf{pseudo-clean targets obtained with:}} \\
\rowcolor[gray]{.9} \cellcolor{white}{} & \multicolumn{2}{c}{} & \multicolumn{2}{c}{\textit{noisereduce}} & \multicolumn{2}{c}{\textit{demucs speech}} & \multicolumn{2}{c|}{\textit{demucs speech+time-scaled}}

\\ \cmidrule(lr){2-9} 
 &  SI-SDR & SI-SDRi & SI-SDR & SI-SDRi & SI-SDR & SI-SDRi & SI-SDR & SI-SDRi & SI-SDR & SI-SDRi 
\\\midrule
small & \SI{6.92}{} $\frac{6.27}{7.57}$ & \SI{2.8}{} $\frac{2.3}{3.35}$ & \SI{12.85}{} $\frac{12.31}{13.45}$ & \SI{9.66}{} $\frac{9.25}{10.25}$ & \SI{12.48}{} $\frac{11.48}{13.33}$ & \SI{8.97}{} $\frac{8.46}{9.37}$ & \SI{12.74}{} $\frac{11.72}{13.6}$ & \SI{9.28}{} $\frac{8.93}{9.75}$ & \SI{8.98}{} $\frac{7.66}{11.16}$ & \SI{5.46}{} $\frac{4.47}{6.57}$ \\\midrule
large & \SI{5.77}{} $\frac{5.58}{5.96}$ & \SI{2.67}{} $\frac{2.58}{2.74}$ & \SI{12.15}{} $\frac{11.98}{12.32}$ & \SI{9.17}{} $\frac{9.07}{9.32}$ & \SI{11.06}{} $\frac{10.9}{11.21}$ & \SI{8.39}{} $\frac{8.23}{8.53}$ & \SI{12.27}{} $\frac{12.08}{12.45}$ & \SI{9.3}{} $\frac{9.15}{9.43}$ & \SI{8.59}{} $\frac{8.35}{8.87}$ & \SI{6.15}{} $\frac{5.74}{6.39}$\\ \midrule
-5dB  & \SI{-4.01}{} $\frac{-4.21}{-3.75}$ & \SI{0.99}{} $\frac{0.78}{1.21}$ & \SI{3.49}{} $\frac{2.45}{4.35}$ & \SI{8.5}{} $\frac{7.47}{9.34}$ & \SI{3.86}{} $\frac{2.96}{4.89}$ & \SI{8.86}{} $\frac{7.95}{9.88}$ & \SI{4.62}{} $\frac{3.55}{5.31}$ & \SI{9.61}{} $\frac{8.53}{10.29}$ & \SI{4.91}{} $\frac{3.39}{6.16}$ & \SI{9.92}{} $\frac{8.37}{11.16}$ \\
0dB   & \SI{1.8}{} $\frac{1.52}{2.04}$ & \SI{1.77}{} $\frac{1.46}{2.02}$ & \SI{10.21}{} $\frac{9.59}{10.65}$ & \SI{10.2}{} $\frac{9.59}{10.62}$ & \SI{9.2}{} $\frac{8.59}{9.96}$ & \SI{9.21}{} $\frac{8.59}{9.96}$ & \SI{10.27}{} $\frac{9.49}{11.02}$ & \SI{10.26}{} $\frac{9.48}{11.0}$ & \SI{7.94}{} $\frac{6.91}{9.29}$ & \SI{7.95}{} $\frac{6.91}{9.29}$ \\
5dB & \SI{8.46}{} $\frac{7.95}{8.85}$ & \SI{3.45}{} $\frac{2.95}{3.85}$ & \SI{14.57}{} $\frac{14.16}{15.02}$ & \SI{9.57}{} $\frac{9.16}{10.02}$ & \SI{13.6}{} $\frac{12.95}{14.04}$ & \SI{8.61}{} $\frac{7.97}{9.04}$ & \SI{14.53}{} $\frac{13.92}{15.16}$ & \SI{9.51}{} $\frac{8.92}{10.15}$ & \SI{9.96}{} $\frac{8.96}{11.74}$ & \SI{4.95}{} $\frac{3.96}{6.74}$ \\
10dB  & \SI{14.78}{} $\frac{14.29}{15.18}$ & \SI{4.78}{} $\frac{4.29}{5.18}$ & \SI{18.28}{} $\frac{17.82}{18.68}$ & \SI{8.28}{} $\frac{7.82}{8.67}$ & \SI{17.27}{} $\frac{16.87}{17.72}$ & \SI{7.26}{} $\frac{6.9}{7.72}$ & \SI{17.88}{} $\frac{17.39}{18.35}$ & \SI{7.88}{} $\frac{7.38}{8.35}$ & \SI{11.72}{} $\frac{10.1}{13.16}$ & \SI{1.72}{} $\frac{0.1}{3.17}$ 
\end{tabular}
\end{table*}

Similarly to speech enhancement, we evaluate our model using an SNR metric, the Scale-Invariant Signal-to-Distortion Ratio (SI-SDR)~\cite{le2019sdr} in \SI{}{dB} units on the benchmarking set in Section \ref{ssec:benchmark}. 
In addition, we compute the improvement of this metric over the noisy mixtures SI-SDRi. 

The metrics are averaged for the full excerpts as follows. For each test excerpt we take the means of the metrics across the seeds. We report the median and the median absolute deviation across 62 files in the test set. 
For the ablation experiments we compute differences at each excerpt between the SI-SDR corresponding to the denoising obtained with the default configuration model and the denoising obtained from the model trained with the ablation condition. 
%
For reproducibility we use a random seed that controls the weight initialization for random conditions and the mixture generation, specifically, the SNR, the time-scaling augmentation, and the pairing between targets and noise. 
For the overall comparison we train models for seeds 0-9, while the ablations are computed for seed $0$.

\subsection{Results and Discussion}
\label{ssec:results}
We compare noisereduce with the demucs dns48 architecture trained with noisy or pseudo-clean targets and we present the results in Table \ref{tab:res1}. We did not include the baseline performance of the speech models because they were sub-par (medians of \SI{-22}{dB} and \SI{-10.4}{dB} for time-scaling on the `small' set). 
We observe that noisereduce produces good quality pseudo-clean targets (\SI{8.98}{dB}) and using them as pseudo-clean targets further improves the denoising ($\approx$12 \SI{}{dB} on `small' and `large'). Surprisingly, using the low-quality estimations from demucs as targets produces results competitive with the higher-quality noisereduced targets ($\approx$12 \SI{}{dB} on `small' and `large'). 
The current neural approach used in bioacoustics denoising, i.e. using noisy targets ~\cite{bergler2020orca}, has the lowest results. These tendencies are further observed on the `large' set, containing all possible combinations of vocalizations and noises, though lower for all target types with the exception of denoised time-scaled targets. 

The SI-SDR values are lower under very noisy, low SNR, conditions, i.e. `-5dB' set. 
Note that the training set mixtures are created for mismatched conditions (0-20 \SI{}{dB} SNR), to test the generalization capabilities of the approaches. 
Given that, when looking at the improvement over all SNR conditions, we observe that SI-SDRi for models trained with pseudo-clean targets approaches fluctuates less ($\approx$ \SI{9}{dB}), proving good generalization and robustness, while the `noisy target' performs poorly on very noisy data.
Another surprising result is that the method `noisereduce' has the most improvement when the signals are more noisy. This may be because noise is more salient and easier to estimate with this kind of approach.

\begin{table}[ht]
    \footnotesize
  \caption{Ablations: Median with 95\% confidence intervals  of \underline{SI-SDR differences} in \SI{}{dB} between the denoising models trained with the default configuration and with the ablation condition across the excerpts in the small test subset for the four training targets types. Positive medians indicate better results for the ablation model.}
  \label{tab:res2}

\begin{tabular}{@{}llll@{}} 
\textbf{noisy} &  \multicolumn{3}{c}{\textbf{pseudo-clean targets}} \\

\textbf{targets} & \textit{noisereduce} & \textit{demucs speech} & \textit{demucs s. time-scaled}
\\\midrule
\rowcolor[gray]{.9} \multicolumn{4}{c}{\textit{Ablation condition: start with random weights}} 
\\
\SI{1.48}{} $\frac{0.4}{2.18}$ & \SI{-1.2}{} $\frac{-2.3}{-0.45}$ & \SI{-2.15}{} $\frac{-2.87}{-0.56}$ & \SI{-1.75}{} $\frac{-2.69}{-0.9}$ \\\midrule
\rowcolor[gray]{.9} \multicolumn{4}{c}{\textit{Ablation condition : CleanUNet instead of demucs dns48}} 
\\
\centering
\SI{-0.56}{} $\frac{-1.27}{-0.03}$ & \SI{-5.37}{} $\frac{-7.84}{-3.62}$ & \SI{-2.15}{} $\frac{-2.87}{-0.56}$ & \SI{-2.43}{} $\frac{-4.22}{-1.29}$
\\\midrule
\rowcolor[gray]{.9} \multicolumn{4}{c}{\textit{Ablation: no random time-scaling augmentation at training}} \\
\SI{0.88}{} $\frac{-0.25}{1.85}$ & \SI{-0.07}{} $\frac{-1.28}{0.28}$ & \SI{-1.09}{} $\frac{-2.34}{0.24}$ & \SI{0.08}{} $\frac{-0.51}{0.49}$ 
\end{tabular}
\end{table}

We perform ablation studies with respect to three conditions. First, we start training with random weights rather than fine-tuning the demucs dns48 model. This improves solely for the noisy targets conditions (\SI{1.48}{dB}) and we conclude that the demucs speech enhancement offers good priors for pseudo-clean target training. 
Second, we replace the demucs model with the CleanUNet to obtain the pseudo-clean targets. We found the pseudo-clean targets obtained with CleanUNet to be of inferior quality. In particular, applying time-scaling to animal signals to lower their pitch gave worse results, i.e. blabber-like speech. When we trained CleanUNet on the demucs time-scaled targets we observed CleanUNet still underperforming (\SI{-2.43}{dB}) even when trained with the same data as demucs. Animal audio signals are more sparse than human speech and it could be that the complex attention mechanisms of CleanUNet leads to overfitting.
Third, we looked at time-scaling as a training augmentation. We found it to only have an impact on performance when training with the demucs pseudo-clean targets. 

\section{Conclusions}
\label{sec:conclusions}



This work proposes a novel technique to train deep learning models for the task of denoising animal vocalizations that does not require the availability of clean, noise-free recordings. In addition, we introduce two non-overlapping data sets for training and benchmarking.
We show that using pre-denoised vocalizations in the form of pseudo-clean targets leads to superior results to the current state of the art noisy-target methods~\cite{bergler2020orca} or `noisereduce'\cite{sainburg2020finding}.
Particularly, we found that `demucs dns48', a model for human speech enhancement, provides effective priors for this task. A similar finding was observed in the context of bioacoustics domain adaptation~\cite{boudiaf2023search,ghani2023global,kloots2024exploring,sarkar2024utility,cauzinilleinvestigating}, suggesting that deep learning models may exploit signal characteristics that are consistent across different taxa and contexts. 
In future work, we plan to investigate the enhancements achievable through subsequent iterations of our method and to extend the model to higher sample rates.

\bibliographystyle{IEEEtran}
\bibliography{main}

\begin{thebibliography}{10}
\providecommand{\url}[1]{#1}
\csname url@samestyle\endcsname
\providecommand{\newblock}{\relax}
\providecommand{\bibinfo}[2]{#2}
\providecommand{\BIBentrySTDinterwordspacing}{\spaceskip=0pt\relax}
\providecommand{\BIBentryALTinterwordstretchfactor}{4}
\providecommand{\BIBentryALTinterwordspacing}{\spaceskip=\fontdimen2\font plus
\BIBentryALTinterwordstretchfactor\fontdimen3\font minus \fontdimen4\font\relax}
\providecommand{\BIBforeignlanguage}[2]{{%
\expandafter\ifx\csname l@#1\endcsname\relax
\typeout{** WARNING: IEEEtran.bst: No hyphenation pattern has been}%
\typeout{** loaded for the language `#1'. Using the pattern for}%
\typeout{** the default language instead.}%
\else
\language=\csname l@#1\endcsname
\fi
#2}}
\providecommand{\BIBdecl}{\relax}
\BIBdecl

\bibitem{fournet2018more}
M.~Fournet, L.~Jacobsen, C.~M. Gabriele, D.~K. Mellinger, and H.~Klinck, ``More of the same: Allopatric humpback whale populations share acoustic repertoire,'' \emph{PeerJ}, vol.~6, p. e5365, 2018.

\bibitem{lin2020source}
T.-H. Lin and Y.~Tsao, ``Source separation in ecoacoustics: A roadmap towards versatile soundscape information retrieval,'' \emph{Remote Sensing in Ecology and Conservation}, vol.~6, no.~3, pp. 236--247, 2020.

\bibitem{denton2022improving}
T.~Denton, S.~Wisdom, and J.~Hershey, ``Improving bird classification with unsupervised sound separation,'' in \emph{ICASSP}.\hskip 1em plus 0.5em minus 0.4em\relax IEEE, 2022, pp. 636--640.

\bibitem{bradbury1998principles}
J.~Bradbury, S.~Vehrencamp \emph{et~al.}, \emph{Principles of animal communication}.\hskip 1em plus 0.5em minus 0.4em\relax Sinauer Associates Sunderland, MA, 1998, vol. 132.

\bibitem{stowell2022computational}
D.~Stowell, ``Computational bioacoustics with deep learning: a review and roadmap,'' \emph{PeerJ}, vol.~10, p. e13152, 2022.

\bibitem{fischer2013bioacoustic}
J.~Fischer, R.~Noser, and K.~Hammerschmidt, ``Bioacoustic field research: a primer to acoustic analyses and playback experiments with primates,'' \emph{American journal of primatology}, vol.~75, no.~7, pp. 643--663, 2013.

\bibitem{defossez2020real}
A.~Defossez, G.~Synnaeve, and Y.~Adi, ``Real time speech enhancement in the waveform domain,'' \emph{arXiv preprint arXiv:2006.12847}, 2020.

\bibitem{kong2022speech}
Z.~Kong, W.~Ping, A.~Dantrey, and B.~Catanzaro, ``Speech denoising in the waveform domain with self-attention,'' in \emph{ICASSP}.\hskip 1em plus 0.5em minus 0.4em\relax IEEE, 2022, pp. 7867--7871.

\bibitem{sainburg2020finding}
T.~Sainburg, M.~Thielk, and T.~Gentner, ``Finding, visualizing, and quantifying latent structure across diverse animal vocal repertoires,'' \emph{PLoS computational biology}, vol.~16, no.~10, p. e1008228, 2020.

\bibitem{panayotov2015librispeech}
V.~Panayotov, G.~Chen, D.~Povey, and S.~Khudanpur, ``Librispeech: an asr corpus based on public domain audio books,'' in \emph{ICASSP}.\hskip 1em plus 0.5em minus 0.4em\relax IEEE, 2015, pp. 5206--5210.

\bibitem{wichern2019wham}
G.~Wichern, J.~Antognini, M.~Flynn, L.~Zhu, E.~McQuinn, D.~Crow, E.~Manilow, and J.~Le~Roux, ``Wham!: Extending speech separation to noisy environments,'' \emph{arXiv preprint arXiv:1907.01160}, 2019.

\bibitem{fonseca2021fsd50k}
E.~Fonseca, X.~Favory, J.~Pons, F.~Font, and X.~Serra, ``Fsd50k: an open dataset of human-labeled sound events,'' \emph{IEEE/ACM Transactions on Audio, Speech, and Language Processing}, vol.~30, pp. 829--852, 2021.

\bibitem{reddy2020interspeech}
C.~Reddy, V.~Gopal, R.~Cutler, E.~Beyrami \emph{et~al.}, ``The interspeech 2020 deep noise suppression challenge: Datasets, subjective testing framework, and challenge results,'' \emph{arXiv preprint arXiv:2005.13981}, 2020.

\bibitem{wisdom2020unsupervised}
S.~Wisdom, E.~Tzinis, H.~Erdogan, R.~Weiss, K.~Wilson, and J.~Hershey, ``Unsupervised sound separation using mixture invariant training,'' \emph{NeurIPS}, vol.~33, pp. 3846--3857, 2020.

\bibitem{ito2023audio}
N.~Ito and M.~Sugiyama, ``Audio signal enhancement with learning from positive and unlabeled data,'' in \emph{ICASSP}.\hskip 1em plus 0.5em minus 0.4em\relax IEEE, 2023, pp. 1--5.

\bibitem{tzinis2022remixit}
E.~Tzinis, Y.~Adi, V.~Ithapu, B.~Xu, P.~Smaragdis, and A.~Kumar, ``Remixit: Continual self-training of speech enhancement models via bootstrapped remixing,'' \emph{IEEE Journal of Selected Topics in Signal Processing}, vol.~16, no.~6, pp. 1329--1341, 2022.

\bibitem{fujimura2023analysis}
T.~Fujimura and T.~Toda, ``Analysis of noisy-target training for dnn-based speech enhancement,'' in \emph{ICASSP}.\hskip 1em plus 0.5em minus 0.4em\relax IEEE, 2023, pp. 1--5.

\bibitem{hao2023neural}
X.~Hao, C.~Xu, and L.~Xie, ``Neural speech enhancement with unsupervised pre-training and mixture training,'' \emph{Neural Networks}, vol. 158, pp. 216--227, 2023.

\bibitem{saijo2023self}
K.~Saijo and T.~Ogawa, ``Self-remixing: Unsupervised speech separation via separation and remixing,'' in \emph{ICASSP}.\hskip 1em plus 0.5em minus 0.4em\relax IEEE, 2023, pp. 1--5.

\bibitem{bergler2020orca}
C.~Bergler, M.~Schmitt, A.~Maier, S.~Smeele, V.~Barth, and E.~Noth, ``Orca-clean: A deep denoising toolkit for killer whale communication,'' in \emph{Interspeech}.\hskip 1em plus 0.5em minus 0.4em\relax ISCA, 2020, pp. 1136--1140.

\bibitem{barnhill2024animal}
A.~Barnhill, E.~Nöth, A.~Maier, and C.~Bergler, ``{ANIMAL}-{CLEAN} – {A} {Deep} {Denoising} {Toolkit} for {Animal}-{Independent} {Signal} {Enhancement},'' in \emph{Interspeech}.\hskip 1em plus 0.5em minus 0.4em\relax ISCA, 2024.

\bibitem{kumar2024vision}
S.~Kumar, J.~Li, and Y.~Zhang, ``Vision transformer segmentation for visual bird sound denoising,'' in \emph{Interspeech}.\hskip 1em plus 0.5em minus 0.4em\relax ISCA, 2024.

\bibitem{stowell2015denoising}
D.~Stowell and R.~Turner, ``Denoising without access to clean data using a partitioned autoencoder,'' \emph{arXiv preprint arXiv:1509.05982}, 2015.

\bibitem{sarkar2024utility}
E.~Sarkar and M.~Doss, ``On the utility of speech and audio foundation models for marmoset call analysis,'' in \emph{VIHAR}, 2024.

\bibitem{cauzinilleinvestigating}
J.~Cauzinille, B.~Favre, D.~Marxer, Rand~Clink, A.~Ahmad, and A.~Rey, ``Investigating self-supervised speech models’ ability to classify animal vocalizations: The case of gibbon’s vocal identity,'' in \emph{Interspeech}.\hskip 1em plus 0.5em minus 0.4em\relax ISCA, 2024.

\bibitem{kloots2024exploring}
M.~de~Heer~Kloots and M.~Knörnschild, ``Exploring bat song syllable representations in self-supervised audio encoders,'' in \emph{VIHAR}, 2024.

\bibitem{ghani2023global}
B.~Ghani, T.~Denton, S.~Kahl, and H.~Klinck, ``Global birdsong embeddings enable superior transfer learning for bioacoustic classification,'' \emph{Scientific Reports}, vol.~13, no.~1, p. 22876, 2023.

\bibitem{code-datasets}
``{Code biodenoising-datasets},'' \url{https://github.com/earthspecies/biodenoising-datasets}, accessed:2024-09-01.

\bibitem{asa}
``{Animal Sound Archive at the Museum für Naturkunde Berlin},'' \url{https://www.museumfuernaturkunde.berlin/en/science/animal-sound-archive}, accessed:2024-01-01.

\bibitem{watkins}
``{Watkins Marine Mammal Sound Database, Woods Hole Oceanographic Institution and the New Bedford Whaling Museum},'' \url{https://whoicf2.whoi.edu/science/B/whalesounds/about.cfm}, accessed:2024-01-01.

\bibitem{akbal2023explainable}
E.~Akbal, P.~Barua, S.~Dogan, T.~Tuncer, and U.~Acharya, ``Explainable automated anuran sound classification using improved one-dimensional local binary pattern and tunable q wavelet transform techniques,'' \emph{Expert Systems with Applications}, vol. 225, p. 120089, 2023.

\bibitem{gustison2012derived}
M.~Gustison, A.~le~Roux, and T.~Bergman, ``Derived vocalizations of geladas (theropithecus gelada) and the evolution of vocal complexity in primates,'' \emph{Philosophical Transactions of the Royal Society B: Biological Sciences}, vol. 367, no. 1597, pp. 1847--1859, 2012.

\bibitem{fukushima2015distributed}
M.~Fukushima, A.~Doyle, M.~Mullarkey, M.~Mishkin, and B.~Averbeck, ``Distributed acoustic cues for caller identity in macaque vocalization,'' \emph{Royal Society open science}, vol.~2, no.~12, p. 150432, 2015.

\bibitem{dufourq2022passive}
E.~Dufourq, C.~Batist, R.~Foquet, and I.~Durbach, ``Passive acoustic monitoring of animal populations with transfer learning,'' \emph{Ecological Informatics}, vol.~70, p. 101688, 2022.

\bibitem{jeantet2023improving}
L.~Jeantet and E.~Dufourq, ``Improving deep learning acoustic classifiers with contextual information for wildlife monitoring,'' \emph{Ecological Informatics}, vol.~77, p. 102256, 2023.

\bibitem{poupard2020massive}
M.~Poupard, P.~Best, M.~Ferrari, P.~Spong, H.~Symonds, J.-M. Pr{\'e}vot, T.~Soriano, and H.~Glotin, ``From massive detections and localisations of orca at orcalab over three years to real-time survey joint to environmental conditions,'' in \emph{e-Forum Acusticum 2020}, 2020, pp. 3235--3237.

\bibitem{ness2013orchive}
S.~Ness, P.~Symonds, Hand~Spong, and G.~Tzanetakis, ``{The Orchive: Data mining a massive bioacoustic archive},'' \emph{arXiv preprint arXiv:1307.0589}, 2013.

\bibitem{glotin2014soundscape}
H.~Glotin, \emph{{Soundscape Semiotics: Localization and Categorization}}.\hskip 1em plus 0.5em minus 0.4em\relax BoD--Books on Demand, 2014.

\bibitem{noaa}
``{NOAA},'' \url{https://www.noaa.gov/}, accessed:2024-03-01.

\bibitem{mars}
``{MBARI MARS},'' \url{https://www.mbari.org/}, accessed:2024-01-01.

\bibitem{santos2016shipsear}
D.~Santos-Dom{\'\i}nguez, S.~Torres-Guijarro, A.~Cardenal-L{\'o}pez, and A.~Pena-Gimenez, ``{ShipsEar: An underwater vessel noise database},'' \emph{Applied Acoustics}, vol. 113, pp. 64--69, 2016.

\bibitem{irfan2021deepship}
M.~Irfan, Z.~Jiangbin, S.~Ali, M.~Iqbal, Z.~Masood, and U.~Hamid, ``{DeepShip: An underwater acoustic benchmark dataset and a separable convolution based autoencoder for classification},'' \emph{Expert Systems with Applications}, vol. 183, p. 115270, 2021.

\bibitem{page}
``{Materials},'' \url{https://mariusmiron.com/research/biodenoising}, accessed:2024-01-01.

\bibitem{freesound}
``{Freesound},'' \url{https://freesound.org/}, accessed: 2024-01-01.

\bibitem{benchmark}
``{Biodenoising\_validation},'' \url{https://zenodo.org/records/13736465}, accessed:2024-01-01.

\bibitem{guzhov2022audioclip}
A.~Guzhov, F.~Raue, J.~Hees, and A.~Dengel, ``Audioclip: Extending clip to image, text and audio,'' in \emph{ICASSP}.\hskip 1em plus 0.5em minus 0.4em\relax IEEE, 2022, pp. 976--980.

\bibitem{code}
``{Code biodenoising},'' \url{https://github.com/earthspecies/biodenoising}, accessed:2024-09-01.

\bibitem{le2019sdr}
J.~Le~Roux, S.~Wisdom, H.~Erdogan, and J.~Hershey, ``Sdr--half-baked or well done?'' in \emph{ICASSP}.\hskip 1em plus 0.5em minus 0.4em\relax IEEE, 2019, pp. 626--630.

\bibitem{boudiaf2023search}
M.~Boudiaf, T.~Denton, B.~van Merri{\"e}nboer, V.~Dumoulin, and E.~Triantafillou, ``In search for a generalizable method for source free domain adaptation,'' \emph{arXiv preprint arXiv:2302.06658}, 2023.

\end{thebibliography}

\end{document}